\documentclass[a4paper,11pt]{amsart}
\begin{document}

\hyphenation{gra-vi-ta-tio-nal re-fe-ren-ces Re-la-ti-vi-ty
ge-ne-ra-ted ma-the-ma-ti-cal ana-ly-ti-cal me-cha-nism
ex-pe-ri-men-tal ge-ne-ral ap-pro-xi-ma-tion po-pu-la-tion
sui-ta-ble re-fe-ren-ce bi-cha-racte-ri-stics phy-si-cal
par-ti-cu-lar con-si-de-rat-tions stra-ta-gems ma-te-ma-ti-che
pro-perty con-si-de-ra-tions pro-pa-ga-tion co-va-riance sy-stem
cha-racte-ri-stics me-moir im-pli-citly rea-li-zed pri-vi-leges
bo-dies }
\title[Relativity and wavy motions]
{{\bf Relativity and wavy motions}}
\author{Angelo Loinger}
\date{}
\address{Dipartimento di Fisica, Universit\`a di Milano, Via
Celoria, 16 - 20133 Milano (Italy)}
\email{angelo.loinger@mi.infn.it}
\date{}
\thanks{To be published on \emph{Spacetime \& Substance}}

\begin{abstract}
The conditions under which the undulatory character of field
disturbances is physically significant.
\end{abstract}

\maketitle

\noindent \emph{\textbf{Summary.}} -- \textbf{1}. Introduction. --
\textbf{2}. and \textbf{3}. Physical meaning of covariance in
general relativity (GR) -- Einstein \emph{v.} Fock -- Refutation
of Fock's viewpoint; consequences concerning the gravitational
waves (GW's). -- \textbf{4}. Electromagnetic waves and GW's. --
\textbf{5}. Characteristics of Einstein and Maxwell fields. --
\textbf{6}. An example of Einsteinian characteristic. --
\textbf{6bis}. Singularities in GR. -- \textbf{7}. The propagation
speed of gravitation. -- \textbf{8}. GW's in the linear
approximation of GR. -- Appendix: Levi-Civita, Einstein and the
GW's. -- Parergon. \vskip1.20cm

\noindent {\bf 1.}-- The theme of the present Note has been
already touched by me in previous papers, \emph{passim}, and
treated expressly in my article ``Waves and uniformity of
space-times" \cite{1}. However, it has not attracted the attention
of the concerned scholars and therefore I give here a little
different treatment of the problem with some further illustration.

\vskip0.80cm
\noindent {\bf 2.}-- The spacetime of \emph{special} relativity
(SR) is usually described by the simple Minkowskian tensor
$\eta_{jk}, (j,k=0,1,2,3)$, for which: $\eta_{rs}=0$ if
\linebreak[5] $r \neq s$, $\eta_{00}=1$,
$\eta_{11}=\eta_{22}=\eta_{33}=-1$; the theory is Lorentz
invariant, the Galilean reference frames are \emph{physically}
privileged. As far back as 1917, it was emphasized by Kretschmann
\cite{2} -- and in subsequent years by Fock \cite{3} -- that
\emph{any} physical theory can be reformulated in \emph{general}
co-ordinates without losing its characteristic physical
properties. In the case of SR, this means in particular that the
Galilean frames maintain their physical privileges; see e.g.
chapt. IV of Fock's treatise cited in \cite{3}. But in
\emph{general} relativity (GR) things stand otherwise: a pregnant
formulation by Erwin Schr\"odinger \cite{4} tells us that: ``The
geometric structure of the space-time model envisaged in the 1915
theory $[$GR$]$ is embodied in the following two principles: (i)
equivalence of all four-dimensional systems of coordinates
obtained from any one of them by arbitrary
(point-)transformations; (ii) the continuum has a metrical
connexion impressed on it: that is, at every point a certain
quadratic form of the coordinate-differentials,
$g_{ik}\mathrm{d}x_{i}\mathrm{d}x_{k}$, called the 'square of the
interval'  between the two points in question, has a fundamental
meaning, invariant in the aforesaid transformations. $[\ldots]$
The first $[$principle$]$, the principle of general invariance,
incarnates the idea of General Relativity." In other terms, in GR
no system of co-ordinates has physical privileges. Seemingly, this
concept has been accepted by the overwhelming majority of
physicists (with the remarkable exception represented by Fock).
However, I wish to emphasize that not all theoreticians have
realized \emph{all} its implications.

\vskip1.20cm
\noindent {\bf 3.}-- Point (i) of the above Schr\"odinger's
quotation means that the phrase ``General Relativity" must be
understood literally: no reference system has physical privileges,
no physically sensible result depends on the chosen co-ordinates.
In any specific instance one chooses of course the reference frame
that is the most adequate and simple for geometric and formal
reasons, avoiding the introduction of superfluous ``inertial
forces" (\emph{lato sensu}), which would complicate uselessly the
computations. The choice of the reference system has only a
practical meaning because in GR the co-ordinates are mere
\emph{labels} for the point-events of spacetime.

\par A first important \emph{consequence}: \emph{no} fundamental
velocity exists in GR, in particular light loses a privilege that
it had in SR: now, its propagation speed depends on the
inertial-gravitational forces, and can vary from zero to infinity.

\par Fock \cite{3} did not share Einstein's standpoint on reference
systems. For him only the \emph{special} theory (SR) deserves the
name of ``theory of relativity". (From a strictly geometrical
viewpoint, ``relativity" and ``uniformity" of the concerned
manifold -- see \emph{infra}, sect.\textbf{4.} -- are closely
related concepts, as it was emphasized in 1927 by E. Cartan.) Fock
tried to demonstrate that in GR -- called by him ``theory of
gravitation" -- there exists a set of infinite reference frames
endowed with an outstanding significance, the \emph{harmonic}
frames. However, his proof rests on unjustified assumptions and
therefore is not a real proof. As a matter of fact, the harmonic
references are useful for the practical solution of many problems,
but do not possess a particular conceptual value.

\par As it is well known, the harmonic co-ordinates, say
$\xi^{j}$, are such that the components $g_{jk}$'s,
$(j,k=0,1,2,3)$, of metric tensor satisfy the following four
conditions $(k=0,1,2,3)$ \cite{5}:

\begin{equation} \label{eq:one}
\frac{\partial}{\partial \xi^{j}} \left(g^{jk}\sqrt{-g}\right)=0
\quad{,}
\end{equation}

where $g:=\det ||g_{jk}||$. According to Fock \cite{5}, the
$\xi$'s are especially appropriate for evidencing in various
stages of approximation the mathematical properties of the GW's.
In Fock's conception, the fact that the undulatory character of
gravitational wavy motions  remains unchanged for all the harmonic
frames is sufficient to make certain the \emph{physical} reality
of the GW's.  Now, two fundamental remarks can be opposed to Fock.
\emph{In primis}, no physical ``mechanism" exists really in the
\emph{\textbf{exact}} GR for the production of GW's; \emph{in
particular}, it is easy to prove that the trajectories of the
bodies of a system, which interact \emph{only} gravitationally,
are \emph{geodesic} lines \cite{6}. Accordingly, even if we
adopted  Fock's conception, the GW's would reveal themselves as
mere analytical artefacts.  \emph{In secundis}, since the correct
interpretation of the formalism of Einstein's field theory affirms
that the phrase ``General Relativity" must be taken \emph{au pied
de la lettre} \cite{4}, we can ascertain immediately that the
undulatory character of any GW is generally impaired by a change
of general co-ordinates. Further, the GW's have only a
\emph{pseudo} stress-energy-momentum tensor.

\vskip1.20cm
\noindent {\bf 4.}-- It is instructive to compare the e.m. waves
with the GW's. The propagation substrate of the e.m. waves of
Maxwell theory is Minkowski spacetime, that is a \emph{uniform}
(i.e., homogeneous and isotropic) manifold, for which the infinite
class of the Galilean frames is physically privileged. When we
re-write Maxwell theory according to the formalism of
Riemann-Einstein spacetime, the ``physicality" of its concepts
remains unchanged, in particular the ``physicality" of the e.m.
waves. On the contrary, the GW's are undulations of the metric
tensor $g_{jk}$, which is the ``substance" of Riemann-Einstein
spacetime, i.e. of a \emph{not} ``fixed" manifold, that does
\emph{not} possess a class of physically privileged reference
systems. Emission ``mechanism" of the e.m. waves can be simply the
acceleration of a charge, whereas the acceleration of a mass does
\emph{not} generate any GW \cite{6}. (For the special case of the
GW's in the \emph{linear} approximation of GR, see sect.\textbf{8}
\emph{infra}.)

\vskip1.20cm
\noindent {\bf 5.}-- \emph{Characteristic hypersurfaces} of
Einstein field equations: in the current literature there is a
physically false interpretation of them.

\par They were first written in 1930 by Levi-Civita \cite{7}, who
gave the  correct interpretation.

\par In SR the differential equations of the characteristics and
of the bicharacteristics of Maxwell field can be written
respectively, in a Hamilton-Jacobi form, as follows:

\begin{equation} \label{eq:two}
H:= \frac{1}{2}\,\eta^{jk}p_{j}p_{k}=0 \quad{,} \qquad
\mathrm{with} \quad p_{j}:=\frac{\partial z(y)}{\partial y^{j}}
\quad{;}
\end{equation}

\begin{equation} \label{eq:three}
\frac{\mathrm{d}p_{j}}{\mathrm{d}\sigma} = - \frac{\partial
H}{\partial y^{j}} \quad{,} \qquad
\frac{\mathrm{d}y^{j}}{\mathrm{d}\sigma} = - \frac{\partial
H}{\partial p_{j}} \quad{,}
\end{equation}

where: $\sigma$ is an auxiliary parameter, $z(y)$ is the function
which defines the characteristic hypersurface $z(y)=0$, and
$\mathrm{d}s^{2}=\eta_{jk}\mathrm{d}y^{j}\mathrm{d}y^{k}$ yields
the Minkowskian interval. Equation $z(y)=0$, $[y\equiv
(y^{0},y^{1},y^{2},y^{3})]$, represents physically the
\textbf{\emph{wave front}} \emph{of an e.m. wave}; the
characteristic lines of (\ref{eq:two}) -- the \emph{\textbf{rays}}
of wave front $z(y)=0$, given by eq.(\ref{eq:three}) -- coincide
with the \textbf{\emph{null}} \emph{geodesics} $\mathrm{d}s=0$.
Thus we see that \emph{SR comprises the} \textbf{\emph{geometric
optics}}. (This could be foreseen because the e.m. theory is a
basic ingredient for defining space and time in SR.)

\par Eqs. (\ref{eq:two}) and (\ref{eq:three}) can be immediately
re-written in a system of \emph{general} co-ordinates
$(x^{0},x^{1},x^{2},x^{3})\equiv x$:

\begin{equation} \label{eq:four}
H:= \frac{1}{2}\,g^{jk}p_{j}p_{k}=0 \quad{,} \qquad \mathrm{with}
\quad p_{j}:=\frac{\partial z(x)}{\partial x^{j}} \quad{;}
\end{equation}

\begin{equation} \label{eq:five}
\frac{\mathrm{d}p_{j}}{\mathrm{d}\sigma} = - \frac{\partial
H}{\partial x^{j}} \quad{,} \qquad
\frac{\mathrm{d}x^{j}}{\mathrm{d}\sigma} = - \frac{\partial
H}{\partial p_{j}} \quad{.}
\end{equation}

Now, as it was demonstrated by Whittaker \cite{8}, eqs.
(\ref{eq:four}) and (\ref{eq:five}) are \emph{formally} identical
to the equations that yield characteristics and bicharacteristics
of \emph{Maxwell} field \emph{in a Riemann-Einstein spacetime}
with metric  tensor $g_{jk}(x)$. And Levi-Civita \cite{7} proved
that eqs. (\ref{eq:four}) and (\ref{eq:five}) give also
characteristics and bicharacteristics of \emph{Einstein} field. A
\emph{not} fortuitous coincidence!

\par \emph{Since GR} \textbf{\emph{comprehends}} \emph{SR}, eqs.
(\ref{eq:four}) and (\ref{eq:five}) have a \textbf{\emph{unique}}
\emph{electromagnetic} interpretation  for Maxwell and Einstein
fields, as it was explicitly emphasized by Levi-Civita \cite{7}:
both SR \emph{and} GR comprise constitutionally the
\textbf{\emph{geometric optics}}.

\par On the contrary, in the current opinion (see e.g. Fock
\cite{3}, sect.\textbf{53}) characteristics and bicharacteristics
of Einstein field equations give wave fronts and rays of GW's: an
interpretation vitiated by the wishful thinking concerning the
real existence of these fictive undulations.

\vskip1.20cm
\noindent {\bf 6.}-- We have seen that \emph{both in SR and in GR}
equation $\mathrm{d}s^{2}=0$ gives the geodesic \emph{null} lines
that represent the \emph{light rays} of \textbf{\emph{geometric
optics}}. The propagation velocity of these rays is \emph{not} a
universal  constant \emph{in GR}, because it depends on the
inertial-gravitational forces as described by potential
$g_{jk}(x)$. A simple example will render evident this \emph{well
known} fact; see also \cite{9}.

\par Let us consider the $\mathrm{d}s^{2}$ of Brillouin's form of
solution of Schwarzschild problem (: to find the gravitational
field generated by a material point of mass $M$, at rest)
\cite{10}:

\begin{equation} \label{eq:six}
\mathrm{d}s^{2}=\frac{r}{r+2m}\,c^{2}\mathrm{d}t^{2} -
\frac{r+2m}{r}\,\mathrm{d}r^{2} -
(r+2m)^{2}(\mathrm{d}\theta^{2}+\sin^{2}\theta\mathrm{d}\varphi^{2})
\quad,
\end{equation}

where $m\equiv GM/c^{2}$. Remark that this form is \emph{maximally
extended} since holds for $r>0$.

\par For $\mathrm{d}\theta=\mathrm{d}\varphi=0$, the condition
$\mathrm{d}s^{2}=0$ implies that

\begin{equation} \label{eq:seven}
\frac{\mathrm{d}r}{c\mathrm{d}t}=\pm \frac{r}{r+2m} \quad;
\end{equation}

if $z:=ct-\psi(r)$ is the function of the characteristic $z=0$, we
have:

\begin{equation} \label{eq:eight}
0=2H=\frac{r+2m}{r} \cdot 1 - \frac{r}{r+2m}
\left(\frac{\mathrm{d}\psi}{\mathrm{d}r}\right)^{2} \quad,
\end{equation}

and therefore:

\begin{equation} \label{eq:nine}
\psi(r) = \pm (r+2m\ln r +\mathrm{const}) \quad;
\end{equation}

thus the characteristic has the equation

\begin{equation} \label{eq:ten}
ct = \pm (r+2m\ln r+\mathrm{const}) \quad,
\end{equation}

from which

\begin{equation} \label{eq:eleven}
\frac{\mathrm{d}r}{c\mathrm{d}t} = - \frac{\partial z}{c\partial
t} / \frac{\partial z}{\partial r}= \pm \frac{r}{r+2m} \quad,
\end{equation}

i.e. again result (\ref{eq:seven}), which can be also re-obtained
by means of eqs. (\ref{eq:five}), written for $(r,ct)$,
$(p_{r},p_{0})$. -- \emph{Q.e.d.}

\par If $m=0$, $\mathrm{d}r/\mathrm{d}t=c$, i.e. the value of SR.
A value which is also obtained for $r\rightarrow\infty$. It is not
a general limiting value because it depends on the chosen frame;
with a convenient transformation of general co-ordinates the light
speed can assume any desired value.

\par It is instructive to compare eq.(\ref{eq:seven}) with eq.(4.2)
of \cite{9} ($\alpha$ in (4.2) coincides with the present
$m$):

\begin{equation} \label{eq:sevenprime}
\frac{\mathrm{d}r}{c\mathrm{d}t}=\pm \frac{r-2m}{r} \quad,
\tag{7$^{\prime}$}
\end{equation}

which is obtained from the \emph{standard} form of solution of
Schwarzschild problem (erroneously called ``by Schwarzschild").
Brillouin's form \cite{6} is diffeomorphic to ``exterior" part
$r>2m$ of standard form: a simple fact that makes clear how the
notion of black hole pertain to a fairy-tale. (The baroque form of
solution by Kruskal and Szekeres is a gift of Barmecide).

\vskip1.20cm
\noindent {\bf 6bis.}-- \emph{The singularities in GR}: they are
classifiable in two mathematical categories: curvature
singularities and co-ordinate singularities. However, from the
\emph{physical} standpoint the only essential distinction is
between singularities characterized by the presence in them of
matter and singularities characterized by the absence in them of
matter. The two classifications do not always coincide. For
instance, the singularity of the Brillouin's form (\ref{eq:six})
and the singularity of the \textbf{\emph{original}}
Schwarzschild's form (which can be obtained, e.g.,  from the
standard form by means of formal substitution $r\rightarrow
[r^{3}+(2m)^{3}]^{1/3}$) are not curvature singularities, but
physical ones. On the contrary, the singularity at $r=0$ of the
standard form is a curvature singularity, but it cannot represent
a material point because is a \emph{space-like} locus; and the
singularity at $r=2m$ of the standard form is a co-ordinate and
non-physical singularity.

\par A widespread ``Vulgate" of GR affirms that if Schwarzschild
problem is solved using co-ordinate-free methods (as orthonormal
bases, \emph{etc.}) the result is necessarily the standard form of
solution. Unfortunately, ``Vulgate"'s procedure is impaired by a
logical fallacy: indeed, these authors have already chosen
initially the above form because they write, first of all, the
simple expression
$r^{2}(\mathrm{d}\theta^{2}+\sin^{2}\theta\mathrm{d}\varphi^{2})$
for the angular part. \emph{Et de hoc satis}.

\vskip1.00cm
\noindent {\bf 7.}-- There is a widespread and erroneous
conviction (see e.g. Fock \cite{3}, p.194) according to which
\emph{in GR} gravitation is propagated with the speed of light
\emph{in vacuo}, i.e. with the speed of light in empty space of
SR.

\par The supporters of this false opinion claim that it follows,
e.g., from eqs. (\ref{eq:four}) and (\ref{eq:five}), when
interpreted as differential equations of wave fronts and rays of
GW's. Now, this is trivially wrong even from the viewpoint of the
believers in the physical existence of GW's, because eqs.
(\ref{eq:four}) and (\ref{eq:five}) -- quite \emph{independently
of their interpretation} -- affirm in reality that the concerned
wave fronts and rays have a propagation velocity that
\emph{depends on the metric tensor} $g_{jk}(x)$, even if this
tensor has the form of a mathematical undulation.

\par The non-existence of physical GW's has the following
consequence: if we displace  a mass, its gravitational field and
the related curvature of the interested manifold \emph{displace
themselves along with the mass}: under this respect Einstein field
and Newton field behave in an identical way \cite{11}.

\par For the GW's in the \emph{linear} approximation of GR, see
sect.\textbf{8}.

\vskip1.00cm
\noindent {\bf 8.}-- The case of the GW's according to
\emph{linear} approximation of GR can be quickly dispatched.
Indeed, it is here sufficient that I recall the decisive argument
of sect.\textbf{4} of a recent paper of mine \cite{12}.

\par As it is well known, in the linearized version of GR one puts
approximately

\begin{equation} \label{eq:twelve}
g_{jk} \approx \eta_{jk} + h_{jk} \quad,
\end{equation}

where $\eta_{jk}$ is the customary Minkowskian tensor and the
$h_{jk}$'s are \emph{small} deviations, that in our case represent
the passage of GW's. The essential point is: $h_{jk}$ \emph{is a
tensor} \textbf{\emph{only}} \emph{under} \textbf{\emph{Lorentz}}
\emph{transformations of Galilean co-ordinates}.

\par Now: $\alpha$) suitable \emph{finite} transformations of
\textbf{\emph{general}} co-ordinates can reduce to
\emph{\textbf{zero}} the undulations $h_{jk}$, just because their
tensorial character is only Lorentzian; $\beta$) for the specially
significant instance of \emph{plane} GW's, remark $\alpha$)
implies the reduction to \emph{\textbf{zero}} of the celebrated
transverse-traceless (TT) GW's; $\gamma$) it is true that \emph{in
Minkowski space-time} the $h_{jk}$--waves are propagated with the
light velocity $c$ \emph{in vacuo}; unfortunately, by virtue of
$\alpha$) and $\beta$), they are only phantom entities.

\par It is regrettable that various physicists insist on publishing
useless considerations and computations on $h_{jk}$--waves
\cite{13}. It is time that astrophysical community desist from
beating the air -- and from squandering the money of the
taxpayers.

\small \vskip0.5cm
\par\hfill {\emph{``Dann zuletzt ist unerl\"a\ss{}lich,}
  \par\hfill \emph{Da\ss{} der Dichter manches hasse;}
  \par\hfill \emph{Was unleidlich ist und h\"a\ss{}lich,}
  \par\hfill \emph{Nicht wie Sch\"ones leben lasse."}
  \par\hfill J.W. v. Goethe

\normalsize

\vskip0.80cm
\newpage
\noindent \emph{\textbf{Appendix: Levi-Civita, Einstein and the
GW's}}

\par \vskip0.10cm
I don't doubt that Levi-Civita's electromagnetic interpretation of
the characteristics of Einstein field \cite{7} was propitiated by
his fundamental memoir of 1917 on GR \cite{14}. Indeed, this paper
ends with two basic remarks: \emph{i}) the proposed pseudo tensor
$t_{jk}$ of stress-energy-momentum of Einstein field, which
satisfies the differential conditions ($T_{jk}$ is the matter
tensor)

\begin{equation} \label{eq:thirteen}
\frac{\partial \,[(t_{j}^{k}+T_{j}^{k})\sqrt{-g}\,]}{\partial
x^{k}}=0 \quad,
\end{equation}

is just a \emph{false} tensor, that can be reduced to zero with a
suitable change of co-ordinates; \emph{ii})  from the geometrical
and analytical standpoint, it is \emph{certain} that the left-hand
side of Einstein equations $[R_{jk}-(1/2)g_{jk}R]$ represents the
\emph{true} stress-energy-momentum tensor of Einstein field -- as
it had been pointed out also by Lorentz \cite{15}.

\par \textbf{\emph{Quite independently}}, point \emph{i}) and point
\emph{ii}) tell us that the GW's (which are solutions of
$R_{jk}=0$) are mere mathematical, not physical, undulations, both
in the \emph{exact} GR and in the \emph{linear} approximation of
it.

\par There was a gentleman's disagreement between Einstein and
Levi-Civita. Against point \emph{ii}) Einstein raised an
intuitive, ``sentimental", objection, which represents an implicit
dissatisfaction with the formal structure of his theory: with the
proposal by Levi-Civita and Lorentz, the total energy of a closed
system is always zero, and the conservation of this value does not
imply the further existence of the physical system under any
whatever form.

\par Point \emph{i}): in Pauli's and in Weyl's treatises
\cite{16} we find the various stratagems (by several authors)
having the aim to prove that under convenient spatio-temporal
\emph{asymptotic conditions} the integrals

\begin{equation} \label{eq:fourteen}
J_{j}:= \int (t_{j}^{0}+ T_{j}^{0})  \sqrt{-g}\,
\mathrm{d}x^{1}\mathrm{d}x^{2}\mathrm{d}x^{3}
\end{equation}

give the conserved \emph{total} four-momentum  of a \emph{closed}
physical system. Now, one can remark that it is conceptually
inappropriate to circumvent a property of the differential
formalism by prescribing \emph{ad hoc} conditions to given
integrals; besides, an \emph{ineffective} procedure, for the
following reason. As Lorentz \cite{15} and Klein \cite{17} pointed
out, there are \emph{other} quantities -- say $w_{jk}$,
\emph{different} from the above $t_{jk}$ --, which satisfy the
\emph{same} conditions, in particular such that

\begin{equation} \label{eq:fifteen}
\frac{\partial \,[(wt_{j}^{k}+ T_{j}^{k})  \sqrt{-g}\,]}{\partial
x^{k}}=0 \quad;
\end{equation}

the $w$'s by Lorentz and Klein depend also on the \emph{second}
derivatives of $g_{jk}$, but this does not violate any physical
principle. Now, the value of the corresponding \emph{total}
four-momentum does \textbf{\emph{not}} coincide with the value
given by eqs. (\ref{eq:fourteen}).

\par Of course, Einstein was perfectly aware that results of the
kind (\ref{eq:fourteen}) are only a provisional way out; and on
the other hand he thought that the entire GR is only a provisional
theory!

\par So far as the GW's are concerned, he was  always doubtful
about their physical existence; a careful reading of his lucid
paper of 1937 with N. Rosen \cite{18} is enlightening. And in his
beautiful booklet \emph{The Meaning of Relativity} no mention is
made of the gravitational waves \cite{19}.

\par A last remark  on these waves. Considering the total failure
of all experimental attempts to reveal them, some physicists have
recently revived -- in private correspondences -- an old
\emph{\textbf{conjecture}}: even if the GW's really existed, it
would be impossible to detect them, because the spatio-temporal
deformation induced by the passage of a GW would interest
\emph{both} the apparatuses, i.e. the resonant bar or the
Michelson interferometer, \emph{and} the devices that register
resp. the bar vibrations or the geodesic deviation of the
suspended mirrors of the interferometer. In this second  case it
is necessary to take into account the interaction of the GW with
the light beams in the interferometric arms \cite{20}, because it
gives a \emph{modification} of the displacement of the
interference fringes generated by the above geodesic deviation.

\vskip0.80cm
\noindent \emph{\textbf{Parergon}}

\par \vskip0.10cm
The readers of Number \emph{27} (Spring 2006) of newsletter
\emph{Matter of Gravity} \cite{21} can see that the
astrophysicists of the ``main stream"  are far from a proper
understanding of the repeated experimental failures for detecting
GW's and BH's.

\par The research brief ``Recent progress in binary black hole
simulations" contains a very peculiar assertion: since the BH's
have singularities which can represent a hard obstacle for
numerical simulations, Pretorius Group ``uses black hole excision,
whereby the black hole interior is removed from the computational
grid. This is justified since the event horizon disconnects the
interior causally from the exterior." Clearly, these scholars have
not realized that the event horizon (the singularity at $r=2m$) is
the essence of the (fictive) notion of BH. BH-excision amounts to
substitute the standard form of solution of Schwarzschild problem
with, e.g., Brillouin's form \cite{10} or Schwarzschild's
(\emph{original}) form, that are diffeomorphic to the exterior
part ($r>2m$) of the standard form -- and are maximally extended.

\par The briefs entitled resp. ``What's new in LIGO" and ``LISA
Pathfinder" describe the continuous refinements of the
apparatuses, and declare implicitly a great optimism about the
results of future searches: the detection of the GW's is now
behind the corner.

\par \emph{Stat pro ratione voluntas}.

\end{document}